\documentclass[12pt]{article}%
\usepackage{amsmath}
\usepackage{amsfonts}
\usepackage{amssymb}
\usepackage{graphicx}%
\providecommand{\U}[1]{\protect\rule{.1in}{.1in}}
\makeatletter
\AtBeginDocument{\@ifpackageloaded{natbib}{\ifNAT@numbers\if@filesw\immediate\write\@auxout{\string\global\string\NAT@numberstrue}\fi\fi}{}}
\makeatother
\begin{document}
\begin{titlepage}
\ \\
\begin{center}
\LARGE
{\bf
Ground-State Entanglement Bound\\
for Quantum Energy Teleportation\\
of General Spin-Chain Models
}
\end{center}
\ \\
\begin{center}
\large{
Masahiro Hotta
}\\
\ \\
\ \\
{\it
Department of Physics, Faculty of Science,\\ Tohoku University,
Sendai 980-8578, Japan\\
hotta@tuhep.phys.tohoku.ac.jp
}
\end{center}
\begin{abstract}
Many-body quantum systems in the ground states have zero-point energy due to the uncertainty relation.
In many cases,
the system in the ground state accompanies spatially-entangled energy density fluctuation via
the noncommutativity of the energy density operators,
though the total energy takes a fixed value, i.e. the lowest eigenvalue of the Hamiltonian.
Quantum energy teleportation (QET) is protocols for extraction of the zero-point energy
out of one subsystem using information of a remote measurement of another subsystem.
From an operational viewpoint of protocol users, QET can be regarded as an effective rapid energy
transportation without breaking all physical laws including causality and local energy conservation.
In the protocols, the ground-state entanglement plays a crucial role. In this paper,
we show analytically for a general class of spin-chain systems
that the entanglement entropy is lower bounded by a positive quadratic function of
the teleported energy between the regions of a QET protocol. This supports a general conjecture
that ground-state entanglement is an evident physical resource for energy transportation in the context of QET.
The result may also deepen our understanding of the energy density fluctuation in condensed matter systems
from a new perspective of quantum information theory.
\end{abstract}
\end{titlepage}

\bigskip

\section{Introduction}

\bigskip

\ 

Many-body quantum systems including quantum fields have zero-point energy of
quantum fluctuation in ground states due to the uncertainty relations.
According to the passivity argument of the ground state \cite{passivity}, an
arbitrary non-trivial local operation on the ground\ state does not cause the
extraction of this energy but leads to the injection of additional energy into
the system by exciting the zero-point fluctuation. This is because the local
operation inevitably yields a different state of the system from the ground
state as the lowest energy state and, thereby, the post-operation state
possesses excitation energy. Therefore, in a fixed region, zero-point energy
is useless for a single experimenter. In quantum field theory, expectation
value of energy density operator in the ground state (vacuum state) is usually
renormalized to zero by subtracting a divergent term corresponding to the
zero-point fluctuation. This expresses that the system in the ground state
represents \textit{nothing} in physics and does not have any useful energy.
However, the zero-point energy of many-body systems indeed becomes available
and can be extracted if two separate experimenters (for example, Alice and
Bob) are able to perform local operations and classical communication
(LOCC)\ for a quantum system that possesses an entangled ground state
\cite{hotta1} \cite{hotta11}. One of the key points of this energy extraction
is a fact that many-body systems in the ground states often accompany
spatially-entangled energy density fluctuation via the noncommutativity of the
energy density operators, though the total energy takes a fixed value, i.e.
the lowest eigenvalue of the Hamiltonian. Thus it is possible to perform in a
spatial region an indirect measurement of the energy density fluctuation in a
separate region by use of the ground-state entanglement. First,\ Alice
performs a local measurement of quantum fluctuation of one subsystem in the
ground state. Because of passivity, her measurement device excites zero-point
fluctuation in her region and injects energy to the system. At the expense of
measurement energy consumption, she obtains information about the quantum
fluctuation and then announces it to Bob, who is in a distant region, with a
light velocity that is much faster than the excitation propagation velocity of
the system. It is of significance to note that the measurement result includes
some information about the zero-point fluctuation of the subsystem in Bob's
region via the ground-state entanglement. Based on the information, Bob can
devise a strategy to suppress the zero-point fluctuation. This enables him to
extract the excess energy out of the subsystem in a local ground state much
before the excitation resulting from the energy injected by Alice reaches
Bob's region. Simultaneously, the suppression of zero-point fluctuation
locally generates a region with a smaller value of energy density than that of
the ground state in the system, which compensates for the energy extraction by
his operation, in accordance with the local energy conservation law. From an
operational viewpoint of the protocol users, the energy injected to the system
in the ground state by Alice can be regarded as input of the protocol and the
energy extracted from the subsystem in the local ground state by Bob as
output. Hence, it is, in effect, energy transportation from Alice to Bob,
though it is a one-time transfer for each entangled state just like the
quantum information transfer by conventional protocols of quantum
teleportation \cite{qt}. Thus, this new protocol is referred to as quantum
energy teleportation (QET). In contrast to the case of QET protocols, by using
the standard protocols of quantum teleportation \cite{qt}, it is impossible to
extract and utilize zero-point energy in its receiver region. In an arbitrary
QET protocol, the amount of energy extracted by Bob is less than that injected
by Alice, and it becomes smaller as the distance between them increases. QET
has not been experimentally verified yet, but a realistic experiment, which
may be achievable with present technology, has been proposed that uses quantum
Hall edge currents \cite{yih}. QET affords not only future development of
quantum technology but also various applications for fundamental physics. For
example, this sheds a new light on entanglement in condensed matter systems
from a viewpoint of local energy density fluctuation. Besides, QET may become
a new available tool of the quantum Maxwell's demons who observe local quantum
fluctuations of an interacting many-body system at the zero temperature and
lead us to an extended paradigm of quantum information thermodynamics. In the
past works about the demons \cite{md}, interactions between subsystems that
the demon watches is assumed to be negligibly small. Thus the ground-state
entanglement has not been taken account of even in the low temparature case.
However, QET enables the demon to perform indirect measurements using the
ground-state entanglement in order to extract work as a new tool. QET also has
a close relation to a local-cooling problem of quantum many-body systems. A
local measurement of zero-point fluctuation in a subsystem is generally
accompanied by energy injection to the system and yields an excited state.
Then a natural question arises. Soon after the energy injection, can we
retrieve all the injected energy using only local operations on the measured
subsystem? The answer is no, and some residual energy is unavoidable in the
system for any local-cooling procedure \cite{hotta1}. The reason for the
residual energy is that the local measurement breaks a part of the
ground-state entanglement and the broken entanglement cannot be restored by
local operations. It turns out that the residual energy is lower bounded by
the total amount of teleported energy via QET by use of the information of the
local measurement \cite{hotta1}. Moreover, QET has been recently applied to
black hole physics and gives a new method for decreasing area of the event
horizon \cite{bh}, just like the Hawking radiation \cite{hr}. Let us imagine
that a measurement of quantum fields outside a massive black hole provides
information about the quantum fluctuation. Because the pre-measurement state
of the quantum fields can be approximated by the usual Minkowski vacuum state
in the flat spacetime, positive-energy wave packets of the fields are
generated during the measurement due to the passivity argument. Assume that
the black hole absorbs the wave packects. Then, very significantly, we are
capable of retrieving a part of the absorbed energy outside the horizon by
QET. Using the measurement information, negative energy wave packects can be
generated outside the horizon by extracting positive energy out of the
zero-point fluctuation of the fields. \ The negative energy of the wave
packects propagates across the event horizon of the black hole and may
pair-annihilate with positive energy of matter previously falling inside the
black hole. Hence this QET process is a phenomenon similar to the spontaneous
emission of Hawking radiation which is often referred to as the energy
tunneling out of black holes \cite{hr} \cite{t}. The energy retrieval yields a
decrease in the horizon area, which is proportional to the entropy of the
black hole. This result may provide a profound suggestion about the origin of
black hole entropy from a viewpoint of information theory. QET is one of
promising tools in physics and will increase its advantage in various fields
of research.

If we do not have any distant-region information via ground-state
entanglement, no energy can be teleported by the QET protocols. It seems very
plausible that the amount of teleported energy is closely related to the
amount of the ground-state entanglement. Thus, the conjecture is possible that
QET with a large amount of teleported energy generally requires a large amount
of the entanglement as an evident physical resource. Interestingly this
conjecture has been partially verified for two specific models. For a
two-qubit model \cite{hotta2}, it has been analytically shown that the amount
of ground-state entanglement breaking by a local measurement of one qubit is
lower bounded by a positive value that is proportional to the maximum amount
of energy teleported from the measured qubit to another qubit. For a harmonic
chain model \cite{yh}, a similar relation between entanglement consumption of
local measurement and amount of teleported energy has been found by numerical
analysis. In this paper, we show analytically, for a general class of
spin-chain models, that the ground-state entanglement entropy is lower bounded
by a positive quadratic function of the teleported energy between the regions
of a QET protocol. This general inequality strongly supports the conjecture
mentioned above. The result may also deepen our understanding of the energy
density fluctuation in condensed matter systems from a new perspective of
quantum information theory. In section 2, a brief review of QET is provided.
In section 3, the entanglement bounds in the context of QET are given and
analyzed. In the last section, summary and discussion are provided.

\section{Formula for Energy Teleported by QET}

\ 

In this section, a brief review of QET is provided. Let us consider a general
model of a spin chain with nearest neighbor interaction. Assume that the
ground state $|g\rangle$ is a pure non-degenerate state. The model is
nonrelativistic, and the excitation propagation velocity of the system is
assumed to be much smaller than the velocity of light. The dimension of the
sub-Hilbert space for each spin is assumed to be finite. The energy density
operator at site $n$ is a Hermitian operator and takes the following form:%

\[
T_{n}=X_{n}-\frac{1}{2}\sum_{l}\left(  g_{n-1/2}^{(l)}Y_{n-1}^{(l)}Y_{n}%
^{(l)}+g_{n+1/2}^{(l)}Y_{n}^{(l)}Y_{n+1}^{(l)}\right)  ,
\]
where $X_{n}$ and $Y_{n}^{(l)}$ are local operators acting on a sub-Hilbert
space at spin site $n$ and $g_{n+1/2}^{(l)}$ are coupling constants. The total
Hamiltonian of the system is given by the total sum of energy density
operators:
\[
H=\sum_{n}T_{n}=\sum_{n}X_{n}-\sum_{n,l}g_{n+1/2}^{(l)}Y_{n}^{(l)}%
Y_{n+1}^{(l)}.
\]
Because we later focus on the difference between pre-operation energy and
post-operation energy, we are able to assume, without changing the physics of
the system, that the expectation value of $T_{n}$ for the ground state
$|g\rangle$ is zero as a useful reference point; that is,
\begin{equation}
\langle g|T_{n}|g\rangle=0, \label{1}%
\end{equation}
when an appropriate constant is subtracted from each $X_{n}$. Because the
energy eigenvalue of the ground state $E_{g}$ is computed as
\[
E_{g}=\langle g|H|g\rangle=\sum_{n}\langle g|T_{n}|g\rangle,
\]
Eq. (\ref{1}) also implies that $E_{g}$ is set to zero by subtracting a
constant from the original Hamiltonian:
\begin{equation}
H|g\rangle=0. \label{2}%
\end{equation}
Thus the Hamiltonian is a non-negative operator:%

\[
H\geq0.
\]
It is worthwhile here to stress that, when $T_{n}$ do not commute with each
other, $T_{n}$ can take negative eigenvalues and, thereby, negative average
values even though the total sum of $T_{n}$, namely the Hamiltonian $H$, is
non-negative. For example, the energy density operator at site $n$ of the
Ising model with transverse external magnetic field $b$ can be naturally
introduced as%

\[
T_{n}=X_{n}-\frac{g}{2}Y_{n-1}Y_{n}-\frac{g}{2}Y_{n}Y_{n+1},
\]
where $g$ is a real Ising coupling constant and the local operators are
defined as
\begin{align*}
X_{n}  &  =b\sigma_{n}^{z}-\varepsilon_{n},\\
Y_{n}  &  =\sigma_{n}^{x},
\end{align*}
with irrelevant real constants $\varepsilon_{n}$. The operator $\sigma_{n}%
^{x}$ ($\sigma_{n}^{z}$) is the $x$($z$)-component of Pauli operator at site
$n$. By using the substitution $\varepsilon_{g}=\sum_{n}\varepsilon_{n}$, the
total Hamiltonian takes the standard form
\[
\sum_{n}T_{n}=b\sum_{n}\sigma_{n}^{z}-g\sum_{n}\sigma_{n}^{x}\sigma_{n+1}%
^{x}-\varepsilon_{g}.
\]
When $\varepsilon_{n}$ is selected properly, Eq. (\ref{1}) and Eq. (\ref{2})
hold. In spite of the non-negativeness of $H$, $T_{n}$ has negative average
values except the cases with very specific values of the ratio $g/b$
\cite{hotta1}.

In the case of the QET protocol, Alice stays at site $n=n_{A}$ and Bob stays
at $n=n_{B}$. Let us assume here that Alice is separated enough from Bob and
the site distance between them satisfies
\begin{equation}
\left\vert n_{A}-n_{B}\right\vert \geq3. \label{71}%
\end{equation}
This condition guarantees local property of their operations in the QET
protocols. In the first step, Alice locally performs a general measurement
(POVM measurement) \cite{nc}. The measurement operator is given by $M_{A}%
(\mu),$ which is a local operator at site $n_{A}$ and satisfies the normality condition%

\begin{equation}
\sum_{\mu}M_{A}^{\dag}(\mu)M_{A}(\mu)=1. \label{91}%
\end{equation}
The POVM of this measurement is written as%
\begin{equation}
\Pi_{A}(\mu)=M_{A}^{\dag}(\mu)M_{A}(\mu). \label{90}%
\end{equation}
After the measurement yielding a result $\mu$, its corresponding
post-measurement state is given by%
\[
|\Psi_{1}(\mu)\rangle=\frac{1}{\sqrt{p_{A}(\mu)}}M_{A}(\mu)|g\rangle,
\]
where $p_{A}(\mu)$ is the emergent probability of $\mu$ and is calculated as
\begin{equation}
p_{A}(\mu)=\langle g|\Pi_{A}(\mu)|g\rangle. \label{100}%
\end{equation}
The average post-measurement state is provided by
\begin{equation}
\rho_{1}=\sum_{\mu}p_{A}(\mu)|\Psi_{1}(\mu)\rangle\langle\Psi_{1}(\mu)|.
\label{84}%
\end{equation}
During the measurement, positive energy ($E_{A}$) is undoubtedly injected into
the system because of passivity \cite{passivity}, and it is evaluated as
\[
E_{A}=\operatorname*{Tr}\left[  H\rho_{1}\right]  -\langle g|H|g\rangle.
\]
In the second step of the protocol, Alice announces the measurement result
$\mu$ to Bob via a classical channel. Because the model is nonrelativistic,
the time duration of communication and time evolution of the system can be
omitted by assuming that the communication speed is the velocity of light.
Thus, the information is received by Bob much before the excitation resulting
from the energy injected by Alice reaches Bob's region. In the third step, Bob
performs a $\mu$-dependent local operation on a spin at $n=n_{B}$ in a local
ground state with zero average energy. The unitary operator is given by
\begin{equation}
U_{B}(\mu)=\exp\left(  -i\theta\left(  \mu\right)  G_{B}\left(  \mu\right)
\right)  , \label{11}%
\end{equation}
where $G_{B}\left(  \mu\right)  $ is a generally $\mu$-dependent Hermitian
local operator at $n=n_{B}$ and $\theta\left(  \mu\right)  $ is a real
constant that is dependent on $\mu$ and is usually fixed so as to maximize
Bob's energy gain from QET. After the operation, the state corresponding to
the result $\mu$ is given by
\begin{equation}
|\Psi_{2}(\mu)\rangle=\frac{1}{\sqrt{p_{A}(\mu)}}U_{B}(\mu)M_{A}(\mu
)|g\rangle\label{81}%
\end{equation}
and the average post-operation state is given by%
\begin{equation}
\rho_{2}=\sum_{\mu}p_{A}(\mu)|\Psi_{2}(\mu)\rangle\langle\Psi_{2}(\mu)|.
\label{82}%
\end{equation}
In a specific setting of $\theta\left(  \mu\right)  ~$and $G_{B}\left(
\mu\right)  $ for QET, the total energy of the system decreases during the
operation. The local energy conservation law ensures that this loss in energy
of the system is equal to Bob's energy gain $(E_{B}>0)$ by virtue of his
operation. Because the average value of energy around Bob is zero before the
operation, he actually extracts positive energy $E_{B}$ out of the subsystem
in a local ground state as \textit{nothing}. Thus, $E_{B}$ is called the
teleported energy in the QET protocol and is evaluated as%
\begin{equation}
E_{B}=\operatorname*{Tr}\left[  H\rho_{1}\right]  -\operatorname*{Tr}\left[
H\rho_{2}\right]  . \label{83}%
\end{equation}
To derive a general formula for $E_{B}$ for models of nearest neighbor
interaction, let us introduce a semilocal Hermitian operator $H_{B}$ as
\[
H_{B}=T_{n_{B}-1}+T_{n_{B}}+T_{n_{B}+1}.
\]
This is the total sum of energy density operators on which a local operation
at site $n=n_{B}$ may have a non-trivial influence. $H_{B}$ can be physically
interpreted as a localized energy operator around Bob's site, satisfying
$\langle g|H_{B}|g\rangle=0$. Due to the locality of Alice's measurement, it
is straightforwardly verified by successive use of Eqs. (\ref{84}), (\ref{71})
and (\ref{91}) that
\begin{equation}
\operatorname*{Tr}\left[  H_{B}\rho_{1}\right]  =0. \label{4}%
\end{equation}
After the operation $U_{B}(\mu)$, it has been proven for the QET protocols
\cite{hotta1} that the average value of $H_{B}$ takes a negative value:%

\begin{equation}
\operatorname*{Tr}\left[  H_{B}\rho_{2}\right]  <0. \label{80}%
\end{equation}
Eq. (\ref{80}) describes that Bob's local operation, which enables him to
extract positive energy from a subsystem with zero energy, simultaneously
generates a region with negative energy density around the subsystem due to
local energy conservation. It is worth to recall that the total energy of the
system remains non-negative even after the emergence of the negative-energy
region. Therefore the amount of energy extracted by Bob does not become larger
than that injected by Alice: $E_{B}\leq E_{A}$. Because $U_{B}(\mu)$ is a
local unitary operation at site $n=n_{B}$, this operation does not affect
energy density of site $\bar{n}$ with $\left\vert \bar{n}-n_{B}\right\vert
\geq2$, i.e.
\begin{equation}
\operatorname*{Tr}\left[  T_{\bar{n}}\rho_{1}\right]  -\operatorname*{Tr}%
\left[  T_{\bar{n}}\rho_{2}\right]  =0 \label{1001}%
\end{equation}
holds for such outside sites by virtue of $\left[  T_{\bar{n}},U_{B}%
(\mu)\right]  =0$ and $U_{B}(\mu)^{\dag}U_{B}(\mu)=I.$ By substituting Eqs.
(\ref{4}) and (\ref{1001}) into Eq. (\ref{83}), the following
energy-conservation relation is directly verified:%
\begin{equation}
E_{B}=-\operatorname*{Tr}\left[  H_{B}\rho_{2}\right]  . \label{1000}%
\end{equation}
This indicates that the sum of the energy gain of Bob and the negative
localized energy at site $n_{B}$ of the system after the operation is equal to
zero, that is, the initial value of the localized energy as it should be.
After simple manipulation by successively substituting Eqs. (\ref{82}),
(\ref{81}) and (\ref{90}) into Eq. (\ref{1000}), it can be proven that $E_{B}$
takes the general form
\begin{equation}
E_{B}=-\sum_{\mu}\langle g|\Pi_{A}(\mu)H_{B}(\mu)|g\rangle, \label{3}%
\end{equation}
where $H_{B}(\mu)$ is defined by
\[
H_{B}(\mu)=U_{B}(\mu)^{\dag}H_{B}U_{B}(\mu)
\]
and $\left[  \Pi_{A}(\mu),H_{B}(\mu)\right]  =0$ holds due to Eq. (\ref{71}).
Eq. (\ref{3}) expresses that teleported energy $E_{B}$ is equal to a sum of
ground-state correlation functions of the local POVM operator $\Pi_{A}(\mu)$
at site $n=n_{A}$ and semilocal operators $H_{B}(\mu)$ at site $n=n_{B}$. If
we have no ground-state entanglement, it is easy by using Eq. (\ref{3}) to
check that $E_{B}$ cannot be positive for any $\theta\left(  \mu\right)
$\ and $G_{B}\left(  \mu\right)  $, as follows. For a non-entangled ground
state (separable ground state) that takes the form%

\begin{equation}
|g\rangle=%
{\displaystyle\prod\limits_{n}}
|g_{n}\rangle, \label{23}%
\end{equation}
by using a local pure state $|g_{n}\rangle$ at site $n$, the two-point
correlation function is reduced to the following product form:
\[
\langle g|\Pi_{A}(\mu)H_{B}(\mu)|g\rangle=\langle g|\Pi_{A}(\mu)|g\rangle
\langle g|H_{B}(\mu)|g\rangle.
\]
Using $\langle g|U_{B}(\mu)^{\dag}T_{\bar{n}}U_{B}(\mu)|g\rangle=0$ with
$\left\vert \bar{n}-n_{B}\right\vert \geq2$, the following relation is
proven:
\begin{equation}
\langle g|H_{B}(\mu)|g\rangle=\langle g|U_{B}(\mu)^{\dag}\left(  \sum
_{n=n_{B}-1}^{n_{B}+1}T_{n}\right)  U_{B}(\mu)|g\rangle=\langle g|U_{B}%
(\mu)^{\dag}HU_{B}(\mu)|g\rangle. \label{17}%
\end{equation}
Thus $\langle g|H_{B}(\mu)|g\rangle$ takes a non-negative value due to the
non-negativity of $H$. Taking account of Eq. (\ref{100}), this result means
that $E_{B}$ has a non-positive value for the non-entangled ground state:
\begin{equation}
E_{B}=-\sum_{\mu}p_{A}(\mu)\langle g|U_{B}(\mu)^{\dag}HU_{B}(\mu)|g\rangle
\leq0. \label{25}%
\end{equation}
However, the situation drastically changes for entangled ground states and
$E_{B}$ can actually take a positive value. In order to grasp the reason why
positive $E_{B}$ is allowed, let us consider, for instance, $U_{B}(\mu)$ with
an infinitesimal value of $\theta\left(  \mu\right)  $. In this case, Eq.
(\ref{3}) can be expanded as
\begin{equation}
E_{B}=\sum_{\mu}\theta\left(  \mu\right)  \langle g|\Pi_{A}(\mu)\dot{G}%
_{B}\left(  \mu\right)  |g\rangle+O\left(  \theta^{2}\right)  , \label{5}%
\end{equation}
where $\dot{G}_{B}\left(  \mu\right)  $ is a semilocal Hermitian operator
around site $n=n_{B}$ defined by%
\[
\dot{G}_{B}\left(  \mu\right)  =i\left[  H_{B},G_{B}\left(  \mu\right)
\right]  .
\]
Since $\left[  \Pi_{A}(\mu),\dot{G}_{B}\left(  \mu\right)  \right]  =0$ is
guaranteed by Eq. (\ref{71}), the correlation function $\langle g|\Pi_{A}%
(\mu)\dot{G}_{B}\left(  \mu\right)  |g\rangle$ takes a real number. For an
entangled ground state $|g\rangle$ satisfying $\langle g|\Pi_{A}(\mu)\dot
{G}_{B}\left(  \mu\right)  |g\rangle\neq0$ for some $\mu$, Eq. (\ref{5})
reveals that $E_{B}$ is capable of taking a positive value:%

\[
E_{B}>0
\]
by appropriately choosing the sign of $\theta\left(  \mu\right)  $ so as to
make $\theta\left(  \mu\right)  \langle g|\Pi_{A}(\mu)\dot{G}_{B}\left(
\mu\right)  |g\rangle$ positive. It should be stressed that rather general
spin-chain models with entangled ground states are able to satisfy the
condition of non-vanishing two-point correlation. Hence a very wide class of
spin-chain models are available for QET with $E_{B}$ positive. In conventional
QET protocols \cite{hotta1}, the sign and magnitude of $\theta\left(
\mu\right)  $ are usually determined in order to maximize the positive value
of $E_{B}$.

Eq. (\ref{3}) directly connects $E_{B}$ with two-point correlation functions.
By definition, the correlation functions provide information about how the
zero-point fluctuation at site $n=n_{A}$ is correlated with that at site
$n=n_{B}$. This correlation is caused unquestionably by the ground-state
entanglement. Thus, it is natural to expect that $E_{B}$ has a nontrivial
relation to ground-state entanglement in general, and this is found to be
true. The next section discusses how the ground-state entanglement entropy
between site $n=n_{A}$ and its complementary region is lower bounded by a
positive quadratic function of teleported energy of a general QET protocol.

Before closing this section, a comment is given on the ground-state
entanglement. Our understanding of many-body quantum entanglement is not
enough yet. We have a lot of entanglement measures, which advantages are
indeed verified in some cases \cite{horodecki}. In the bipartite entanglement
case with an energy-sender subsystem $A$ and an energy-receiver subsystem $B$
of QET, several entanglement measures including the negativity and the
log-negativity can be explicitly computed from a reduced density operator
$\rho_{AB}$ of the subsystems. It is known that, even for many-body systems at
criticality at zero temperature, such a bipartite entanglement measure is
calculated as a product of a power law and an exponential decay in terms of
the separation between $A$ and $B$ \cite{qe} \cite{yh}. Thus the bipartite
entanglement, that would be a resource of QET, becomes negligibly small for a
large separation beyond a typical length of the system. However, the amount of
teleported energy from $A$ to $B$ just obeys a power-law decay as the distance
becomes large \cite{hotta1} \cite{bh}.\ Thus the long-distance QET remains
effective even though the bipartite entanglement is not available. This
superficial paradox is resolved by noting that the mutual information between
$A$ and $B$ decays in a power law manner in contrast to the bipartitle
entanglement. QET can be performed only by use of this mutual information
shared by $A$ and $B$. The bipartite entanglement of the two subsystems is not
necessary. However, it should be stressed that this correlation of $A$ and $B$
described by the mutual information is actually generated by not only the
bipartite entanglement but also multipartite entanglements in the ground state
\cite{hotta4}. If the ground state is an exactly separable (non-entangled)
state which takes a product form of pure states of all subsystems, we do not
have such a correlation between them at all. Thus it is a quite natural
attempt to introduce a notion of `the mother entanglement', which gives birth
to the mutual information between $A$ and $B$ for QET. Then what are the most
appropriate entanglement measure for the description of this mother
entanglement? Unfortunately, this remains a serious open problem. However it
can be said, at least, that the entanglement entropy $S_{ent}\left(  A,\bar
{A}\right)  $ of $A$ and its complement $\bar{A}$, which includes $B$,
\ precisely captures the mother entanglement property. In fact, if
$S_{ent}\left(  A,\bar{A}\right)  =0$, no mutual information of $A$ and $B$ is
generated and QET does not work at all. In this sense, this entanglement
entropy is truly a resource of QET for the ground-state case. Therefore, in
the next section, we adopt $S_{ent}\left(  A,\bar{A}\right)  $ in order to
describe how much entanglement the spin-chain systems possess as a QET
resource at zero temperature. At the end a remark is appended for finite
temperature cases. It has been shown very recently that not quantum
entanglement but quantum discord \cite{discord} becomes a resource for
protocols of finite-temperature QET \cite{frey-hotta}. For the Ising spin
model composed of two qubits in the presence of transverse magnetic field, we
have a critical temperature above which entanglement between the qubits in the
thermal state completely vanishes, though the quantum discord (thermal
discord) remains nonzero. By utilizing information shared via the quantum
discord, a high-temperature QET protocol for the two qubits can extract more
energy out of one qubit in the thermal state than that extracted only by use
of local operations.

\bigskip

\bigskip

\section{Ground-State Entanglement Bound in Terms of Teleported Energy}

\ 

\ 

Let us consider a ground state $|g\rangle$ of a general spin-chain model. Let
region $A$ be composed of a single site $n=n_{A}$ and region $B$ be composed
of three sites with $n=n_{B}-1,n_{B},n_{B}+1$ satisfying Eq. (\ref{71}). The
reduced state for $A$ is given by $\rho_{A}=\operatorname*{Tr}_{\bar{A}%
}\left[  |g\rangle\langle g|\right]  $, that for $B$ by $\rho_{B}%
=\operatorname*{Tr}_{\bar{B}}\left[  |g\rangle\langle g|\right]  ,$\ and that
for $A\cup B$ by $\rho_{AB}=\operatorname*{Tr}_{\overline{A\cup B}}\left[
|g\rangle\langle g|\right]  $, where the bar for a set means complement of the
set. Herein, we analytically show, for a general spin-chain model, that
entanglement entropy between $A$ and $\bar{A}$ of a ground state is lower
bounded by a positive quadratic function of energy $E_{B}$ teleported from $A$
to $B$ in an arbitrary QET protocol. To derive the inequality, we first focus
on not entanglement entropy itself but instead mutual information $I_{A:B}$
between $A$ and $B$ defined as
\begin{equation}
I_{A:B}=S_{A}+S_{B}-S_{AB}, \label{6}%
\end{equation}
where $S_{A}=S(\rho_{A})$, $S_{B}=S(\rho_{B})$, $S_{AB}=S(\rho_{AB}),$ and
$S\left(  \rho\right)  $ is the von Neumann entropy of $\rho$:%
\[
S\left(  \rho\right)  =-\operatorname*{Tr}\left[  \rho\ln\rho\right]  .
\]
When $|g\rangle$ is an entangled state, the mutual information $I_{A:B}$ may
take a positive value. \ It is first noted that the following inequality is
proven: For $I_{A:B}$ in Eq. (\ref{6}) and $E_{B}$ in Eq. (\ref{3}),%

\begin{equation}
I_{A:B}\geq\frac{\left\vert E_{B}+\left\langle H\right\rangle \right\vert
^{2}}{2\left\Vert H_{B}\right\Vert ^{2}}, \label{7}%
\end{equation}
where $\left\langle H\right\rangle $ is defined by
\begin{equation}
\left\langle H\right\rangle =\sum_{\mu}p_{A}(\mu)\langle g|U_{B}(\mu)^{\dag
}HU_{B}(\mu)|g\rangle. \label{15}%
\end{equation}
$\left\langle H\right\rangle $ can be interpreted as excitation energy of the
system after performing a probabilistic operation $U_{B}(\mu)$ with its
probability $p_{A}(\mu)$ to the ground state. Unless $U_{B}(\mu)=I$ for each
$\mu$, $\left\langle H\right\rangle $ must take a positive value owing to the
passivity of the ground state. $\left\Vert H_{B}\right\Vert $ in Eq. (\ref{7})
stands for the matrix norm of $H_{B}$ given by the maximum absolute value of
the eigenvalue of $H_{B}$:
\[
\left\Vert H_{B}\right\Vert =\max\left\{  \left\vert \varepsilon
_{B}\right\vert :H_{B}|\varepsilon_{B}\rangle=\varepsilon_{B}|\varepsilon
_{B}\rangle\right\}  .
\]
The proof of Eq. (\ref{7}) is as follows: Let us think a pointer system
$A^{\prime}$ of Alice's measurement device. Consider a complete orthogonal
vector basis $\left\{  |\mu_{A^{\prime}}\rangle:\langle\mu_{A^{\prime}}%
|\mu_{A^{\prime}}^{\prime}\rangle=\delta_{\mu\mu^{\prime}}\right\}  \,$\ in a
Hilbert space of $A^{\prime}$ corresponding to the measurement output
$\left\{  \mu\right\}  $ of measurement operator $M_{A}(\mu)$. Before the
measurement, assume that the pointer state is in a pure state $|0_{A^{\prime}%
}\rangle$. The total initial state of the composite system of $A^{\prime}$,
$A$, and $B$ before the measurement of Alice is given by
\[
\Phi_{A^{\prime}AB}=|0_{A^{\prime}}\rangle\langle0_{A^{\prime}}|\otimes
\rho_{AB}.
\]
Because $|0_{A^{\prime}}\rangle$ is a pure state and $A^{\prime}$ has no
correlation with $A$ and $B$, mutual information $I_{A^{\prime}A:B}$ between
$A^{\prime}\cup A$ and $B$ of $\Phi_{A^{\prime}AB}$ is equal to $I_{A:B}$ of
$|g\rangle\langle g|$ between \thinspace$A$ and $B$:
\[
I_{A^{\prime}A:B}=I_{A:B}.
\]
Let us consider a quantum operation $\Gamma$ for $\Phi_{A^{\prime}AB}$ that
describes the measurement of Alice and satisfies
\[
\Gamma\left[  \Phi_{A^{\prime}AB}\right]  =\sum_{\mu}|\mu_{A^{\prime}}%
\rangle\langle\mu_{A^{\prime}}|\otimes M_{A}(\mu)\rho_{AB}M_{A}(\mu)^{\dag}.
\]
After performing the operation, we discard subsystem $A$ and define a reduced
state $\rho_{A^{\prime}B}$ defined by
\[
\rho_{A^{\prime}B}=\operatorname*{Tr}_{A}\left[  \Gamma\left[  \Phi
_{A^{\prime}AB}\right]  \right]  .
\]
Note that $\rho_{B}=\operatorname*{Tr}_{A^{\prime}}\left[  \rho_{A^{\prime}%
B}\right]  $ holds because of the locality of Alice's measurement. Taking
account of this relation, the mutual information $I_{A^{\prime}:B}$ between
$A^{\prime}$ and $B$ after the manipulation is computed as
\[
I_{A^{\prime}:B}=S(\rho_{A^{\prime}})+S(\rho_{B})-S(\rho_{A^{\prime}B}),
\]
where%
\begin{equation}
\rho_{A^{\prime}}=\sum_{\mu}p_{A}(\mu)|\mu_{A^{\prime}}\rangle\langle
\mu_{A^{\prime}}|, \label{8}%
\end{equation}%
\begin{equation}
\rho_{A^{\prime}B}=\sum_{\mu}|\mu_{A^{\prime}}\rangle\langle\mu_{A^{\prime}%
}|\otimes\operatorname*{Tr}_{A}\left[  \Pi_{A}(\mu)\rho_{AB}\right]  .
\label{9}%
\end{equation}
Let us define a $\mu$-dependent post-measurement state $\rho_{B}(\mu)$ as
\begin{equation}
\rho_{B}(\mu)=\frac{1}{p_{A}(\mu)}\operatorname*{Tr}_{A}\left[  \Pi_{A}%
(\mu)\rho_{AB}\right]  =\frac{1}{p_{A}(\mu)}\operatorname*{Tr}_{\bar{B}%
}\left[  \Pi_{A}(\mu)|g\rangle\langle g|\right]  . \label{14}%
\end{equation}
Then, we are able to rewrite Eq. (\ref{9}) in a transparent form that
describes the perfect correlation between $A^{\prime}$ and $B$ of the
post-measurement state as%
\begin{equation}
\rho_{A^{\prime}B}=\sum_{\mu}p_{A}(\mu)|\mu_{A^{\prime}}\rangle\langle
\mu_{A^{\prime}}|\otimes\rho_{B}(\mu). \label{10}%
\end{equation}
It is a well-known monotonicity property that both quantum operation and
discard of subsystems never increase mutual information \cite{nc}. This
monotonicity can be proven by use of strong subadditivity of the von Neumann
entropy \cite{LR}. Therefore, the following relation holds:
\[
I_{A^{\prime}A:B}\geq I_{A^{\prime}:B}.
\]
Because $I_{A:B}=I_{A^{\prime}A:B}$, this implies the following inequality:
\begin{equation}
I_{A:B}\geq I_{A^{\prime}:B}. \label{18}%
\end{equation}
Here, it is worthwhile to note a useful relation of relative entropy \cite{OP}
that
\begin{equation}
S(\rho||\varphi)\geq\frac{1}{2}\left(  \left\Vert \rho-\varphi\right\Vert
_{1}\right)  ^{2}, \label{31}%
\end{equation}
where $S(\rho||\varphi)=\operatorname*{Tr}\left[  \rho\ln\rho\right]
-\operatorname*{Tr}\left[  \rho\ln\varphi\right]  $ for two quantum states
$\rho$ and $\varphi$, and $\left\Vert \rho-\varphi\right\Vert _{1}$ is the
trace norm of $\rho-\varphi$ given by $\left\Vert \rho-\varphi\right\Vert
_{1}=\operatorname*{Tr}\left[  \sqrt{\left(  \rho-\varphi\right)  ^{2}%
}\right]  $. The proof of Eq. (\ref{31}) is outlined in Appendix 1. Since
$I_{A^{\prime}:B}$ is expressed by using relative entropy as
\[
I_{A^{\prime}:B}=S(\rho_{A^{\prime}B}||\rho_{A^{\prime}}\rho_{B}),
\]
the following inequality is satisfied:%

\[
I_{A^{\prime}:B}=S(\rho_{A^{\prime}B}||\rho_{A^{\prime}}\rho_{B})\geq\frac
{1}{2}\left(  \left\Vert \rho_{A^{\prime}B}-\rho_{A^{\prime}}\rho
_{B}\right\Vert _{1}\right)  ^{2}.
\]
Because%

\begin{equation}
\left\Vert X\right\Vert _{1}\geq\frac{\left\vert \operatorname*{Tr}\left[
XY\right]  \right\vert }{\left\Vert Y\right\Vert } \label{30}%
\end{equation}
holds for $\,$arbitrary Hermitian operators $X$ and $Y$ as proven in Appendix 2,%

\[
\frac{1}{2}\left(  \left\Vert \rho_{A^{\prime}B}-\rho_{A^{\prime}}\rho
_{B}\right\Vert _{1}\right)  ^{2}\geq\frac{\left\vert \operatorname*{Tr}%
\left[  \rho_{A^{\prime}B}M_{A^{\prime}B}\right]  -\operatorname*{Tr}\left[
\rho_{A^{\prime}}\rho_{B}M_{A^{\prime}B}\right]  \right\vert ^{2}}{2\left\Vert
M_{A^{\prime}B}\right\Vert ^{2}}%
\]
holds for an arbitrary Hermitian operator $M_{A^{\prime}B}$ \ of the composite
system of $A^{\prime}$ and $B$. This inequality is provided by Wolf et al
\cite{wvhc} in local operator product cases: $M_{A^{\prime}B}=M_{A^{\prime}%
}\otimes M_{B}$. In later discussion, we fix $M_{A^{\prime}B}$ in a specific
form by use of Bob's operation $U_{B}(\mu)=\exp\left(  -i\theta\left(
\mu\right)  G_{B}\left(  \mu\right)  \right)  $. \ Let us introduce a
non-local unitary operator
\[
U_{A^{\prime}B}=\exp\left(  -i\sum_{\mu}\theta(\mu)|\mu_{A^{\prime}}%
\rangle\langle\mu_{A^{\prime}}|G_{B}\left(  \mu\right)  \right)  ,
\]
This operators satisfy%
\begin{equation}
U_{A^{\prime}B}|\mu_{A^{\prime}}\rangle=U_{B}(\mu)|\mu_{A^{\prime}}%
\rangle.\label{101}%
\end{equation}
By using $U_{A^{\prime}B}$, the operator $M_{A^{\prime}B}$ is defined as
follows:
\begin{equation}
M_{A^{\prime}B}=U_{A^{\prime}B}^{\dag}H_{B}U_{A^{\prime}B}.\label{13}%
\end{equation}
Because $\left\Vert M_{A^{\prime}B}\right\Vert =\left\Vert H_{B}\right\Vert $,
we are able to derive the following inequality:%

\[
I_{A^{\prime}:B}\geq\frac{\left\vert \operatorname*{Tr}\left[  \rho
_{A^{\prime}B}U_{A^{\prime}B}^{\dag}H_{B}U_{A^{\prime}B}\right]
-\operatorname*{Tr}\left[  \rho_{A^{\prime}}\rho_{B}U_{A^{\prime}B}^{\dag
}H_{B}U_{A^{\prime}B}\right]  \right\vert ^{2}}{2\left\Vert H_{B}\right\Vert
^{2}}.
\]
Using Eqs. (\ref{10}), (\ref{101}), (\ref{14}), (\ref{3}) successively, it can
be shown that $\operatorname*{Tr}\left[  \rho_{A^{\prime}B}U_{A^{\prime}%
B}^{\dag}H_{B}U_{A^{\prime}B}\right]  $ is equal to $-E_{B}$ as follows:
\begin{align*}
&  \operatorname*{Tr}\left[  \rho_{A^{\prime}B}U_{A^{\prime}B}^{\dag}%
H_{B}U_{A^{\prime}B}\right] \\
&  =\sum_{\mu}p_{A}(\mu)\operatorname*{Tr}_{B}\left[  \rho_{B}(\mu)\langle
\mu_{A^{\prime}}|U_{A^{\prime}B}^{\dag}H_{B}U_{A^{\prime}B}|\mu_{A^{\prime}%
}\rangle\right] \\
&  =\sum_{\mu}p_{A}(\mu)\operatorname*{Tr}_{B}\left[  \rho_{B}(\mu)U_{B}%
(\mu)^{\dag}H_{B}U_{B}(\mu)\right] \\
&  =\sum_{\mu}\operatorname*{Tr}_{B}\left[  \operatorname*{Tr}_{\bar{B}%
}\left[  \Pi_{A}(\mu)|g\rangle\langle g|\right]  U_{B}(\mu)^{\dag}H_{B}%
U_{B}(\mu)\right] \\
&  =-E_{B}.
\end{align*}
Similarly, using Eqs. (\ref{8}), (\ref{101}), (\ref{17}) and (\ref{15}), it
can be proven that $\operatorname*{Tr}\left[  \rho_{A^{\prime}}\rho
_{B}U_{A^{\prime}B}^{\dag}H_{B}U_{A^{\prime}B}\right]  $ is equal to
$\left\langle H\right\rangle $ as follows:
\begin{align*}
&  \operatorname*{Tr}\left[  \rho_{A^{\prime}}\rho_{B}U_{A^{\prime}B}^{\dag
}H_{B}U_{A^{\prime}B}\right] \\
&  =\sum_{\mu}p_{A}(\mu)\langle g|U_{B}(\mu)^{\dag}H_{B}U_{B}(\mu)|g\rangle\\
&  =\sum_{\mu}p_{A}(\mu)\langle g|U_{B}(\mu)^{\dag}HU_{B}(\mu)|g\rangle\\
&  =\left\langle H\right\rangle .
\end{align*}
Therefore, we obtain the following inequality:%

\[
I_{A^{\prime}:B}\geq\frac{\left\vert E_{B}+\left\langle H\right\rangle
\right\vert ^{2}}{2\left\Vert H_{B}\right\Vert ^{2}},
\]
and Eq. (\ref{7}) is proven because of Eq. (\ref{18}). The result
simultaneously yields another inequality:%
\[
I_{A:B}\geq\frac{E_{B}{}^{2}}{2\left\Vert H_{B}\right\Vert ^{2}}.
\]
This implies that performing the QET with teleported energy $E_{B}$ requires
the mutual information more than $E_{B}{}^{2}/(2\left\Vert H_{B}\right\Vert
^{2})$. Thus it can be said that the mutual information $I_{A:B}$ is a
resource of QET. However, as emphasized in the last paragraph of section 2,
$I_{A:B}$ is generated by the ground-state entanglement. Therefore it is quite
natural to rewrite the result in terms of the entanglement entropy
$S_{ent}\left(  A,\bar{A}\right)  $. Since $\bar{A}\supset B$ and the
monotonicity of mutual information holds in discarding subsystems of no
interest \cite{nc},
\begin{equation}
I_{A:\bar{A}}=S\left(  \rho_{A}\right)  +S\left(  \rho_{\bar{A}}\right)
-S\left(  |g\rangle\langle g|\right)  \geq I_{A:B}, \label{20}%
\end{equation}
where $\rho_{\bar{A}}=\operatorname*{Tr}_{A}\left[  |g\rangle\langle
g|\right]  $, is also satisfied. Owing to purity of the ground state,
$S\left(  |g\rangle\langle g|\right)  =0$ and $S\left(  \rho_{A}\right)
=S\left(  \rho_{\bar{A}}\right)  $. Thus, Eq. (\ref{20}) yields the following
inequality:
\begin{equation}
S_{ent}\left(  A,\bar{A}\right)  \geq\frac{1}{2}I_{A:B}, \label{21}%
\end{equation}
where $S_{ent}\left(  A,\bar{A}\right)  $ is given by
\[
S_{ent}\left(  A,\bar{A}\right)  =S\left(  \rho_{A}\right)  =\frac{1}%
{2}I_{A:\bar{A}}.
\]
From Eqs. (\ref{21}) and (\ref{7}), we finally obtain one of our main results:%

\begin{equation}
S_{ent}\left(  A,\bar{A}\right)  \geq\frac{\left\vert E_{B}+\left\langle
H\right\rangle \right\vert ^{2}}{4\left\Vert H_{B}\right\Vert ^{2}}.
\label{22}%
\end{equation}
The equality of Eq. (\ref{22}) is attained for spin-chain models with
separable ground states with its form in Eq. (\ref{23}) because $S_{ent}%
\left(  A,\bar{A}\right)  =0\,\ $and $E_{B}+\left\langle H\right\rangle =0$ as
seen in Eq. (\ref{25}). Eq. (\ref{22}) gives a lower bound for the
ground-state entanglement entropy $S_{ent}\left(  A,\bar{A}\right)  $ for
general spin-chain QET protocols with $E_{B}>0$. Trivially, the following
inequality holds for arbitrary QET protocols since both $E_{B}$ and
$\left\langle H\right\rangle $ in Eq. (\ref{22}) are positive:
\begin{equation}
S_{ent}\left(  A,\bar{A}\right)  \geq\frac{E_{B}^{2}}{4\left\Vert
H_{B}\right\Vert ^{2}}. \label{70}%
\end{equation}
If we know the value of $E_{B}$ in a specific QET protocol, Eq. (\ref{70})
provides a lower bound of ground-state entanglement entropy from an
operational viewpoint. The result of Eq. (\ref{70}) strongly supports the
conjecture that a large amount of ground-state entanglement $S_{ent}\left(
A,\bar{A}\right)  $ is required as an evident physical resource to perform a
QET protocol when the amount of teleported energy\thinspace$E_{B}$ is large.

It is a rather straightforward extension to consider larger regions for energy
sender $A$ and receiver $B$. Let $A$ be Alice's region with $n=n_{A}-l_{A}\sim
n_{A}+l_{A}$ and $B$ be Bob's region with $n=n_{B}-l_{B}\sim n_{B}+l_{B}$ by
setting $l_{A}$ and $l_{B}$ to be positive integers and by assuming that
$\left\vert n_{A}-n_{B}\right\vert \geq2+l_{A}+l_{B}$. Now $M_{A}(\mu)$ is a
measurement operator acting on a composite Hilbert subspace of spins in $A$,
and $U_{B}(\mu)$ is a unitary operator acting on a composite Hilbert subspace
of spins with $n=n_{B}-l_{B}+1\sim n_{B}+l_{B}-1$. Then, $H_{B}$ is redefined
as
\[
H_{B}=\sum_{n=n_{B}-l_{B}}^{n_{B}+l_{B}}T_{n}.
\]
Even after such an extension, our results in Eqs. (\ref{3}), (\ref{15}),
(\ref{22}), and (\ref{70}) still hold. This extension may deepen our
understanding of QET and ground-state entanglement itself. For example, it
turns out that Eq. (\ref{70}) provides a universal upper bound of the ratio
$E_{B}/\left\Vert H_{B}\right\Vert $ independent of the number $2l_{B}-1$ of
energy extraction points of $B$ and the detail of QET protocols. For an
arbitrary fixed subsystem $A$, $E_{B}$ may be expected to become large as
$l_{B}$ becomes large, but $\left\Vert H_{B}\right\Vert $ becomes larger as
well and the ratio $E_{B}/\left\Vert H_{B}\right\Vert $ never exceeds
$2S_{ent}\left(  A,\bar{A}\right)  ^{1/2}$. Note that $E_{B}/\left\Vert
H_{B}\right\Vert \leq1$ holds by definition of the matrix norm. Thus the upper
bound $2S_{ent}\left(  A,\bar{A}\right)  ^{1/2}$ provides valuable information
about the teleported energy when $S_{ent}\left(  A,\bar{A}\right)  <1/4$. The
extension also allows us to treat $A~$and $\bar{A}$ symmetrically even if the
sizes are different. We are able to exchange the roles of $A~$and $\bar{A}$ in
the QET protocols: Now $\bar{A}$ is the measured system, and a subsystem
$\tilde{A}$ of $A$ belonging to $\left[  n_{A}-l_{A}+1,n_{A}+l_{A}-1\right]  $
is the controlled system dependent on the measurement result, out of which
teleported energy $E_{\tilde{A}}$ is extracted. Under such an exchange, the
entanglement entropy remains unchanged because of the ground-state purity:%
\[
S_{ent}\left(  \bar{A},A\right)  =S_{ent}\left(  A,\bar{A}\right)  .
\]
The left hand side of Eq. (\ref{70}) is also unchanged. Therefore the
following inequality holds:%

\[
S_{ent}\left(  A,\bar{A}\right)  \geq\frac{E_{\tilde{A}}^{2}}{4\left\Vert
H_{\tilde{A}}\right\Vert ^{2}},
\]
where $H_{\tilde{A}}$ denotes the localized energy operator of $\tilde{A}$.
This provides another lower bound of the ground-state entanglement entropy.

\bigskip

\bigskip

\section{Summary and Discussion}

\bigskip

\ 

We considered the ground state of a general model of a spin chain with nearest
neighbor interaction to analyze an arbitrary QET protocol. A universal
inequality in Eq. (\ref{22}) is proven. This inequality implies that
ground-state entanglement $S_{ent}(A,\bar{A})$ between the energy sender's
region $A$ and its complementary region $\bar{A}$, which includes the energy
receiver's region $B$, is lower bounded by a positive quadratic function of
teleported energy $E_{B}$. The obtained results still hold even in the
extended settings for large $A$ and $B$. The result of Eq. (\ref{70}), derived
from Eq. (\ref{22}), strongly supports the general conjecture that a large
amount of ground-state entanglement is required as an evident physical
resource to perform a QET protocol when the amount of teleported energy is large.

The results in this paper are expected to deepen our understanding of the
energy density fluctuation in condensed matter systems from a new perspective
of quantum information theory. The teleported energy originally emerges from
the zero-point fluctuation in the ground state. Hence the amount of the energy
and its distance dependence may lead us to fundamental relations of the
condensed matter systems, which can be naturally termed `the
fluctuation-information relation'. For instance, if a state tomography of the
ground state is experimentally achieved, we are able to evaluate the
entanglement entropy $S_{ent}\left(  A,\bar{A}\right)  $. Then the inequality
of Eq. (\ref{70}) yields a upper bound of $E_{B}$ that is closely related to
the amplitude of energy density fluctuation in the ground state.

In this paper, we consider one-dimensional spin chain models. The extension to
higher-dimensional lattice models with nearest neighbor interaction can be
achieved easily and the same results are obtained, especially the result of
Eq. (\ref{70}) holds. As stressed in \cite{wvhc}, the area law of entanglement
entropy $S_{ent}\left(  A,\bar{A}\right)  $ of a compact region $A$ and its
complement $\bar{A}$ in terms of the boundary area is generally proven for the
ground states of the models, and interestingly alludes to relations between
the entanglement entropy and the holographic principle of black hole physics.
When we perform a measurement of $A$ and, by a QET protocol, extract the
corresponding teleported energy $E_{B}$ from an outside region $B$ that almost
overlaps $\bar{A}$ except a buffer area shared with $A$, a upper bound of the
teleported energy
\[
E_{B}\leq2\left\Vert H_{B}\right\Vert S_{ent}\left(  A,\bar{A}\right)  ^{1/2}%
\]
is derived from Eq. (\ref{70}). This suggests a nontrivial area dependence of
$E_{B}$. The analyses of the dependence are of much interest for black hole
physics and may provide a more profound insight into the holographic
principle. The results will be reported elsewhere.

\bigskip

\textbf{Acknowledgments}

I would like to thank Holger F. Hofmann for giving me valuable comments about
the previous manuscript. This research has been partially supported by the
Global COE Program of MEXT, Japan, and the Ministry of Education, Science,
Sports and Culture, Japan, under Grant No. 21244007.

\bigskip

{\LARGE Appendix 1}

{\LARGE \bigskip}

\ 

In this appendix, a proof outline of Eq. (\ref{31}) in \cite{OP}\thinspace\ is
shown. As preparation, let us first consider a function $f_{x}(y)$ of
\thinspace$x$ and $y$ \ with $0\leq x\leq y\leq1$ defined by
\[
f_{x}(y)=x\ln\left(  \frac{x}{y}\right)  +\left(  1-x\right)  \ln\left(
\frac{1-x}{1-y}\right)  -2\left(  y-x\right)  ^{2}.
\]
The partial derivative in terms of $y$ is found to be non-negative:%

\[
\partial_{y}f_{x}(y)=\frac{\left(  1-2y\right)  ^{2}}{y(1-y)}\left(
y-x\right)  \geq0.
\]
Therefore, the minimum value of $f_{x}(y)$ in terms of $y$ for a fixed value
of $x$ is zero:
\[
\min_{x\leq y\leq1}f_{x}(y)=f_{x}(x)=0.
\]
Hence, $f_{x}(y)$ is non-negative and the inequality
\begin{equation}
2\left(  x-y\right)  ^{2}\leq x\ln\left(  \frac{x}{y}\right)  +\left(
1-x\right)  \ln\left(  \frac{1-x}{1-y}\right)  \label{50}%
\end{equation}
holds for $0\leq x\leq y\leq1$. Next, let us consider the spectral
decomposition of $\rho-\varphi$, where $\rho$ and $\varphi$ are density
operators of quantum states:%

\[
\rho-\varphi=\sum_{\lambda}\lambda P(\lambda).
\]
Here, $\lambda$ is an eigenvalue of $\rho-\varphi$ and $P(\lambda)$ is its
corresponding projective operator. Let us define a projective operator $P_{+}$
by a sum of $P(\lambda)$ with non-negative eigenvalues:
\[
P_{+}=\sum_{\lambda\geq0}P(\lambda).
\]
Further, let us introduce a projection operator $P_{-}$ as the complement of
$P_{+}$:
\[
P_{-}=\sum_{\lambda<0}P(\lambda)=I-P_{+}.
\]
We define emergent probabilities of the ideal measurement result of $P_{\pm}$
for the two states $\rho$ and $\varphi$ as follows:
\begin{align*}
p_{\pm}  &  =\operatorname*{Tr}\left[  \rho P_{\pm}\right]  ,\\
q_{\pm}  &  =\operatorname*{Tr}\left[  \varphi P_{\pm}\right]  .
\end{align*}
Then, the trace norm $\left\Vert \rho-\varphi\right\Vert _{1}$ is computed as
\begin{align*}
\left\Vert \rho-\varphi\right\Vert _{1}  &  =\sum_{\lambda}\left\vert
\lambda\right\vert =\sum_{\lambda\geq0}\lambda-\sum_{\lambda<0}\lambda\\
&  =\operatorname*{Tr}\left[  \left(  \rho-\varphi\right)  P_{+}\right]
-\operatorname*{Tr}\left[  \left(  \rho-\varphi\right)  P_{-}\right] \\
&  =2\operatorname*{Tr}\left[  \left(  \rho-\varphi\right)  P_{+}\right] \\
&  =2\left(  p_{+}-q_{+}\right)  .
\end{align*}
Because the inequality
\[
2\left(  p_{+}-q_{+}\right)  ^{2}\leq p_{+}\ln\left(  \frac{p_{+}}{q_{+}%
}\right)  +p_{-}\ln\left(  \frac{p_{-}}{q_{-}}\right)
\]
is generally satisfied according to Eq. (\ref{50}), the following relation
holds:
\[
\frac{1}{2}\left(  \left\Vert \rho-\varphi\right\Vert _{1}\right)
^{2}=2\left(  p_{+}-q_{+}\right)  ^{2}\leq p_{+}\ln\left(  \frac{p_{+}}{q_{+}%
}\right)  +p_{-}\ln\left(  \frac{p_{-}}{q_{-}}\right)  =S_{c}(p||q),
\]
where $S_{c}(p||q)$ is the classical relative entropy of $p$ and $q$. Because
of the monotonicity of the relative entropy \cite{OP}, the classical relative
entropy $S_{c}(p||q)$ is upper bounded by the quantum relative entropy
$S(\rho||\varphi)$:%
\[
S_{c}(p||q)\leq S(\rho||\varphi).
\]
Therefore, we obtain Eq. (\ref{31}).

\bigskip

\bigskip

\bigskip

$\bigskip$

{\LARGE Appendix 2}

\bigskip

\ 

In this appendix, we prove a standard triangular inequality of the trace norm
and the matrix norm in Eq. (\ref{30}). Let $X$ and $Y\,$\ be Hermitian
operators. Let us introduce a spectral decomposition of $X$ and $Y$ as%
\begin{equation}
X=\sum_{n}x_{n}|x_{n}\rangle\langle x_{n}|, \label{51}%
\end{equation}%
\begin{equation}
Y=\sum_{m}y_{m}|y_{m}\rangle\langle y_{m}|. \label{52}%
\end{equation}
Then, the matrix norm of $Y$ is written as%
\[
\left\Vert Y\right\Vert =\max_{m}\left\vert y_{m}\right\vert =\left\vert
y\right\vert _{\max}.
\]
Eqs. (\ref{51}) and (\ref{52}) yield the following relation:%

\begin{align}
\frac{\left\vert \operatorname*{Tr}\left[  XY\right]  \right\vert }{\left\Vert
Y\right\Vert }  &  =\left\vert \sum_{n}x_{n}\sum_{m}\frac{y_{m}}{\left\vert
y\right\vert _{\max}}\left\vert \langle x_{n}|y_{m}\rangle\right\vert
^{2}\right\vert \nonumber\\
&  \leq\sum_{n}\left\vert x_{n}\right\vert \left\vert \sum_{m}\frac{y_{m}%
}{\left\vert y\right\vert _{\max}}\left\vert \langle x_{n}|y_{m}%
\rangle\right\vert ^{2}\right\vert . \label{54}%
\end{align}
Because
\[
0\leq\frac{\left\vert y_{m}\right\vert }{\left\vert y\right\vert _{\max}}%
\leq1
\]
holds, the right-hand-side term in Eq. (\ref{54}) is upper bounded by
$\left\Vert X\right\Vert _{1}$ as follows:
\begin{align*}
&  \sum_{n}\left\vert x_{n}\right\vert \left\vert \sum_{m}\frac{y_{m}%
}{\left\vert y\right\vert _{\max}}\left\vert \langle x_{n}|y_{m}%
\rangle\right\vert ^{2}\right\vert \\
&  \leq\sum_{n}\left\vert x_{n}\right\vert \left\vert \sum_{m}\left\vert
\langle x_{n}|y_{m}\rangle\right\vert ^{2}\right\vert \\
&  =\sum_{n}\left\vert x_{n}\right\vert =\left\Vert X\right\Vert _{1}.
\end{align*}
Thus, we obtain Eq. (\ref{30}).

\end{document}